\documentclass[runningheads]{llncs}
\usepackage[T1]{fontenc}
\usepackage{graphicx}
\usepackage{amsmath}
\usepackage{amsfonts} 
\begin{document}
\title{A simple strategy for constructing ultradiscrete systems that exhibit the same dynamics as a given continuous model}
\titlerunning{Cellular automata with prescribed dynamics}
\author{Ralph Willox %\inst{1}
\orcidID{0000-0002-8054-0239} %\and
%Second Author\inst{2,3}\orcidID{1111-2222-3333-4444} \and
%Third Author\inst{3}\orcidID{2222--3333-4444-5555}
}
\authorrunning{R. Willox}
\institute{Graduate School of Mathematical Sciences, the University of Tokyo, 3-8-1 Komaba, Meguro-ku, Tokyo, 153-8914 Japan\\
\email{willox@ms.u-tokyo.ac.jp}\\
}
\maketitle              % typeset the header of the contribution
\begin{abstract}
Over the past 30 years, the special limiting procedure known as {\em ultradiscretisation} has become the tool of choice in the field of infinite dimensional integrable systems for constructing cellular automata that exhibit solitonic behaviour, as e.g. in the Korteweg-de Vries equation. A lesser known fact is that, in many cases, ultradiscretisation can also be used to construct cellular automata that retain the essential dynamical features (such as the existence of limit cycles etc.) of a dynamical system expressed in terms of ordinary differential equations (ODEs). In its standard application, the ultradiscretisation procedure relies on the prior construction of a `good' discretisation of the dynamical system at hand, that shares the essential dynamical features of the ODE and which is sign-free, making it amenable to the ultradiscrete limit. In this paper we show, on a simple but generic model, that even if one starts from a discrete system with different dynamical features than the ODE one obtains as its continuum limit,  the ultradiscrete limit can be tweaked such that the dynamics of the resulting cellular automaton is closer to that of the continuum limit than to that of the discrete model itself.

\end{abstract}
\section{Introduction}
The {\em ultradiscretisation} procedure was introduced in \cite{To1996} to explain the solitonic nature of the famous Box\&Ball system (BBS) \cite{TS1990} . The discovery that a special singular limit, performed on a well-chosen representation of an integrable discrete analogue of the Korteweg-de Vries (KdV) equation, could be used not only to construct the BBS from the discrete KdV equation but also to obtain explicit representations of the solitons in the BBS, has led to a wave of activity on ultradiscrete integrable models (see e.g. \cite{IKT2012} for a survey of its many ramifications in the theory of integrable lattice models, representation theory and tropical geometry). 

In \cite{Wi2003} a first attempt was made to use the ultradiscretisation procedure as a tool to construct cellular automata that possess the same dynamical features as a given ordinary differential equation (ODE). A crucial element in this procedure is to identify a discrete analogue of the ODE at hand, suited to the ultradiscrete limit which can only be performed on non-negative quantities and on equations that only have non-negative coefficients. For first order ODEs of the form
\begin{equation}
\frac{d x}{dt} = x\,  \left(F(x) - G(x)\right),\label{xpeq}
\end{equation}
where $^\forall x\geq0, F(x), G(x)\geq0$, such a discretisation (with $t=n \tau$, for some time step $\tau>0$) has been proposed in \cite{Al2005,Ca2006}, in the form of the first order map
\begin{equation}
x_{n+1} =\frac{1 + \tau \big( F(x_n) + \Delta(x,\tau) \big)}{1 + \tau \big( G(x_n) + \tilde\Delta(x,\tau) \big)}\,  x_n ,\label{genmap}
\end{equation}
which is free of minus signs and therefore amenable to ultradiscretisation. ($\Delta(x,\tau)$ and $\tilde\Delta(x,\tau)$ are essentially free, non-negative, functions of $x$ and $\tau$, only constrained by $\lim_{\tau\to+0} \Delta(x,\tau) = \lim_{\tau\to+0} \tilde\Delta(x,\tau)$.)
For example, for the well-known logistic equation $(r,h>0)$
\begin{equation}
\dfrac{d x}{dt} = x (r - h x),\label{logisticeq}
\end{equation}
for $\Delta(x,\tau)= \frac{e^{r\tau}-1-r\tau}{\tau},\, \tilde\Delta(x,\tau)=\frac{h}{r} x \Delta(x,\tau)$, one obtains
\begin{equation}
x_{n+1} = \frac{e^{r\tau} x_n}{1 + \frac{h}{r} (e^{r\tau}-1) x_n},\label{dlogistic}
\end{equation}
which, as a matter of fact, describes an exact sampling \cite{Mo1965} of the solution of \eqref{logisticeq} for $x(0)=x_0\in (0,\frac{r}{h}]$:
\begin{equation}
x_n=\frac{r x_0}{h x_0 + (r- h x_0) e^{-r\tau n}}.\label{exactsample}
\end{equation}
 Note that $e^{r\tau}-1>0$ and that the above recurrence relation is indeed sign-free. The ultradiscrete limit of \eqref{dlogistic} is then obtained by setting
$\tau=\frac{1}{\epsilon}\,,~ x_n=\dfrac{r}{h} e^{-X_n/\epsilon}$,
for $\epsilon, X_n >0$ (so as to respect the constraint $0<x_n<r/h$) and then taking the limit $\epsilon\to+0$ of both sides of the equation, which yields
\begin{equation}
X_{n+1} = X_n - r + \max[0, r-X_n].\label{udlogistic}
\end{equation}
This piece-wise linear map provides a very simple cellular automaton\footnote{Contrary to the BBS for which the ultradiscretisation procedure was introduced, the system described by the ultradiscrete equation \eqref{udlogistic} has no spatial extent and is, strictly speaking, just a 1 dimensional dynamical system over a max-plus algebra. However, as the `state'-variable $X_n$ (even for $r, X_n\in \mathbb{R}_{\geq0}$)  can only take a finite number of values (at most $1+\lfloor X_0/r\rfloor$) it is customary in this field to think of the system as describing something akin to a zero-radius (zero-range) cellular automaton.}, e.g. over the non-negative integers ($r, X\in\mathbb{Z}_{\geq0}$),  that perfectly encapsulates the dynamics of the logistic equation \eqref{logisticeq}, with the crucial difference however that the `asymptotic' value of $X_n$, $X_\infty=0$, is now attained in a \emph{finite} number of steps. This is clearly reflected in the solution of \eqref{udlogistic}, $X_n= \max[0, X_0-n r]$, which itself can also be obtained as the ultradiscrete limit of \eqref{exactsample}.

\section{Ultradiscrete models that exhibit the same dynamics as a given continuous system}
The above construction of a cellular automaton version of the logistic equation depended crucially on the discretisation \eqref{dlogistic}, which is in a sense almost too perfect in that it describes a genuine sampling of an orbit of the logistic equation, rather than a discrete approximation of  it. However, the above example was only chosen for pedagogical expediency and the method works also remarkably well on more general (sign-free) discretisations of ODE-based models that arise in ecology, epidemiology, population dynamics and many other domains, each time yielding (zero-radius) cellular automata versions of the original model that retain all essential dynamics of the original continuous model [4, 6, 8--11].\footnote{The method has also been extended to models that are defined in terms of partial differential equations, see e.g. \cite{Ma2015}.} For ODE-based models some attempts have been made at explaining why the ultradiscretisation procedure is able to preserve the essential dynamical features of the original ODE, in terms of dynamical systems theory \cite{Ya2023}. 

This is not to say, however, that the method is guaranteed to work on any ODE. One notable exception being the 2D Lotka-Volterra system, for which no faithful sign-free discretisation is known; all attempts so far have resulted in discrete chaotic maps, far removed from the simple integrable behaviour of the 2D Lotka-Volterra system (cf. \cite{FFP2024}). Subsequent ultradiscretisation preserves the chaotic nature of the discrete map and hence, all cellular automata obtained in this way necessarily exhibit chaotic behaviour as well \cite{Hi1997}.
This then begs the question whether one always needs a `faithful' discretisation of a given ODE in order to obtain a cellular automaton that will exhibit the dynamical behaviour of that ODE. Here we give an example of an ODE for which its sign-free discretisation exhibits dynamical behaviour that is not present in the ODE, but for which a special ultradiscrete limit eliminates this spurious dynamics, yielding a cellular automaton with the same dynamical features as the original ODE. We believe this to be the first example of its kind.

Let us consider the following extension \cite{Wi2009}  of the Lotka-Volterra system in which the evolution of the prey population $x(t)$ is constrained by both predation (i.e. by the predator population $y(t)$) as well as by the environment (as in the logistic equation):
\begin{equation}
\left\{
\begin{array}{l}
x'= x (1 - x - y)\\
y'= y (\gamma x- \delta),
\end{array}
\right.\label{LVext}
\end{equation}
 ($\delta,\gamma>0$). The dynamics of this system is extremely simple.
The fixed point at the origin $(0,0)$ is a saddle point and the fixed point at $(1,0)$ is a sink when the parameters satisfy $\gamma<\delta$. When $\gamma>\delta$ this fixed point becomes a saddle and a third fixed point appears, $(\frac{\delta}{\gamma}, \frac{\gamma-\delta}{\gamma})$, which is either a sink or a spiral sink.

\begin{figure}[h]
\begin{center}
\includegraphics[width=11cm]{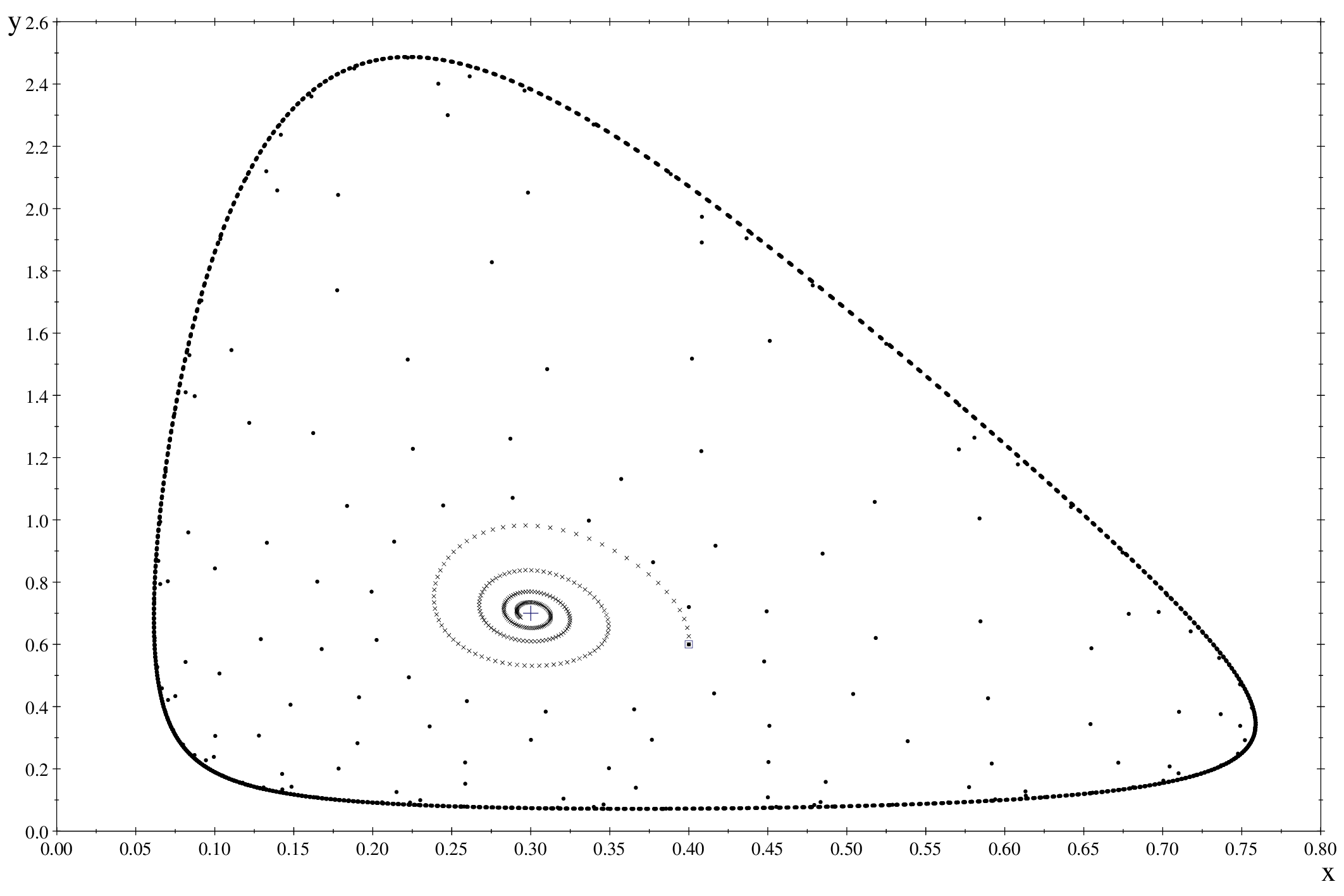}
\caption{Figure showing two orbits for the sign-free discretisation \eqref{LVextNS} of the extended Lotka-Volterra system \eqref{LVext}, for two values of the discretisation step, with $\gamma=5, \delta=1.5$. For a stepsize of the order of 1 the orbit (here shown for 2500 iterations) is rapidly attracted to a limit cycle, but when the stepsize decreases the cycle disappears and the orbit (here shown for 250 iterations at $\tau=10^{-1}$) is attracted to a stable fixed point. The initial condition is the same for both orbits and is indicated by a square box. The fixed point (which is independent of the stepsize) is indicated by a cross.}\label{fig1} 
\end{center}
\end{figure}

The simplest sign-free discretisation of \eqref{LVext} is obtained by extending the ansatz \eqref{genmap} to two dimensions:
\begin{equation}
\left\{
\begin{array}{l}
x_{n+1}= \dfrac{(1+\tau) x_n }{1+ \tau (x_n + y_n)}\\[-1mm] \\
y_{n+1}= \dfrac{(1+\tau\gamma x_n) y_n}{1+\tau\delta}.
\end{array}
\right.\label{LVextNS}
\end{equation}
This 2D map has fixed points at $(0,0)$ and $(1,0)$ that have exactly the same properties as in the continuous case, for all values of $\gamma$ and $\delta$. The third fixed point $(\frac{\delta}{\gamma}, \frac{\gamma-\delta}{\gamma})$ again only exists when $\gamma>\delta$, but when $\gamma>2\delta +\frac{1}{\tau}$ it ceases to be a sink and becomes repulsive. Since the solution of \eqref{LVextNS} cannot move out to infinity it has a limit cycle. Fig.~\ref{fig1} shows two orbits of \eqref{LVextNS}, for $\gamma=5, \delta=1.5$ and the initial condition $(x_0,y_0)=(0.4,0.6)$, one for $\tau=10^{-1}$ for which the condition for the (third) fixed  point $(0.3,0.7)$ to be unstable is violated and the orbit spirals into the fixed point, and one for $\tau=1$ for which that fixed point is unstable and the orbit converges rapidly to a limit cycle. 

The appearance of a spurious limit cycle in the discrete system \eqref{LVextNS} of course bears the hallmark of a typical stepsize-induced bifurcation \cite{Al2005,Oh2023} for large values of $\tau$, at fixed $\gamma$ and $\delta$, which is a common occurrence when discretising a given ODE. However, one can also look at it the opposite way: no matter how small a value one takes for $\tau$, there always exists a parameter range in $\gamma$ and $\delta$ for which the discrete system \eqref{LVextNS} has a dynamical feature (the limit cycle) that the continuous system does not possess (the case $\tau=0$ begin obviously trivial as a discrete system). Looked at it this way we are forced to admit that, generally speaking, \eqref{LVextNS} is {\em not} a faithful discretisation of \eqref{LVext}, and one might therefore assume that any subsequent ultradiscretisation of \eqref{LVextNS} will inherit this problem. We will now show that this is not necessarily the case.

We take the ultradiscrete limit of \eqref{LVextNS} by setting $\tau=e^{T/\epsilon}, \gamma=e^{\Gamma/\epsilon}, \tau \delta=e^{\Delta/\epsilon}, \tau x= e^{X/\epsilon}$ and $\tau y= e^{Y/\epsilon}$, and by taking the limit $\epsilon\to+0$ of both equations. This yields the ultradiscrete system
\begin{equation}
\left\{
\begin{array}{l}
X_{n+1} = \max[0,T] + X_n - \max[0,X_n,Y_n]\\[-1mm] \\
Y_{n+1}= Y_n+\max[0,\Gamma+X_n]-\max[0,\Delta],
\end{array}
\right.\label{LVextud}
\end{equation}
which, in all generality, we shall consider as a cellular automaton over the rational numbers: $T\in\mathbb{Q}\setminus\{0\}, \Gamma, \Delta\in\mathbb{Q}$, and $X, Y\in\mathbb{Q}\cup\{-\infty\}$. \footnote{As remarked before in connection to the ultradiscrete system \eqref{udlogistic}, by this we really mean a `zero-radius' cellular automaton, i.e. without spatial extent or interaction. Generally speaking, unlike \eqref{udlogistic}, this system has an infinite phase-space but as the state variables  $X_n$ and $Y_n$ only change by integer multiples of the number $1/D$, where $D$ is the l.c.m. of the denominators in $X_0,Y_0,\Gamma$ and $\Delta$, it is customary in the field to refer to such a system as a cellular automaton as well.} Given the piece-wise linear character of these equations, there are only two essentially different realizations of this cellular automaton: one for $T=1$ and  one for $T=-1$ in which the first term in the rhs. of the equation for $X_{n+1}$ disappears. It turns out that the former describes the discrete evolution \eqref{LVextNS} in which there are either two fixed points one of which is a sink, or three unstable fixed points resulting in a limit cycle. The latter, on the other hand, exhibits the dynamical behaviour of the continuous system \eqref{LVext} in which there can either be two fixed points (a saddle and a sink) or two unstable fixed points and a sink. The system for $T=-1$ is actually very simple and one easily obtains its general behaviour: the fixed point $(X_*,Y_*)=(-\infty,-\infty)$ is always repulsive; for $\Delta>0$ generic initial data $(X_0,Y_0)$ are attracted to a fixed point $(X_*,-\infty)$ for some finite $X_*<0$, except when $X_0=\Delta-\Gamma\leq0, Y_0\leq0$ which is a fixed point of a third, different, type or, in case $\Delta=\Gamma$, for which $X_0>0, Y_0\leq-X_0$ is attracted to $(0,X_0+Y_0)$; for $\Delta\leq0$ any initial condition is also attracted to a fixed point of this third type: $(X_*,Y_*)$ with $Y_*\neq-\infty$ and $X_*\leq0$.

\begin{figure}[h]
\begin{center}
\includegraphics[width=11cm]{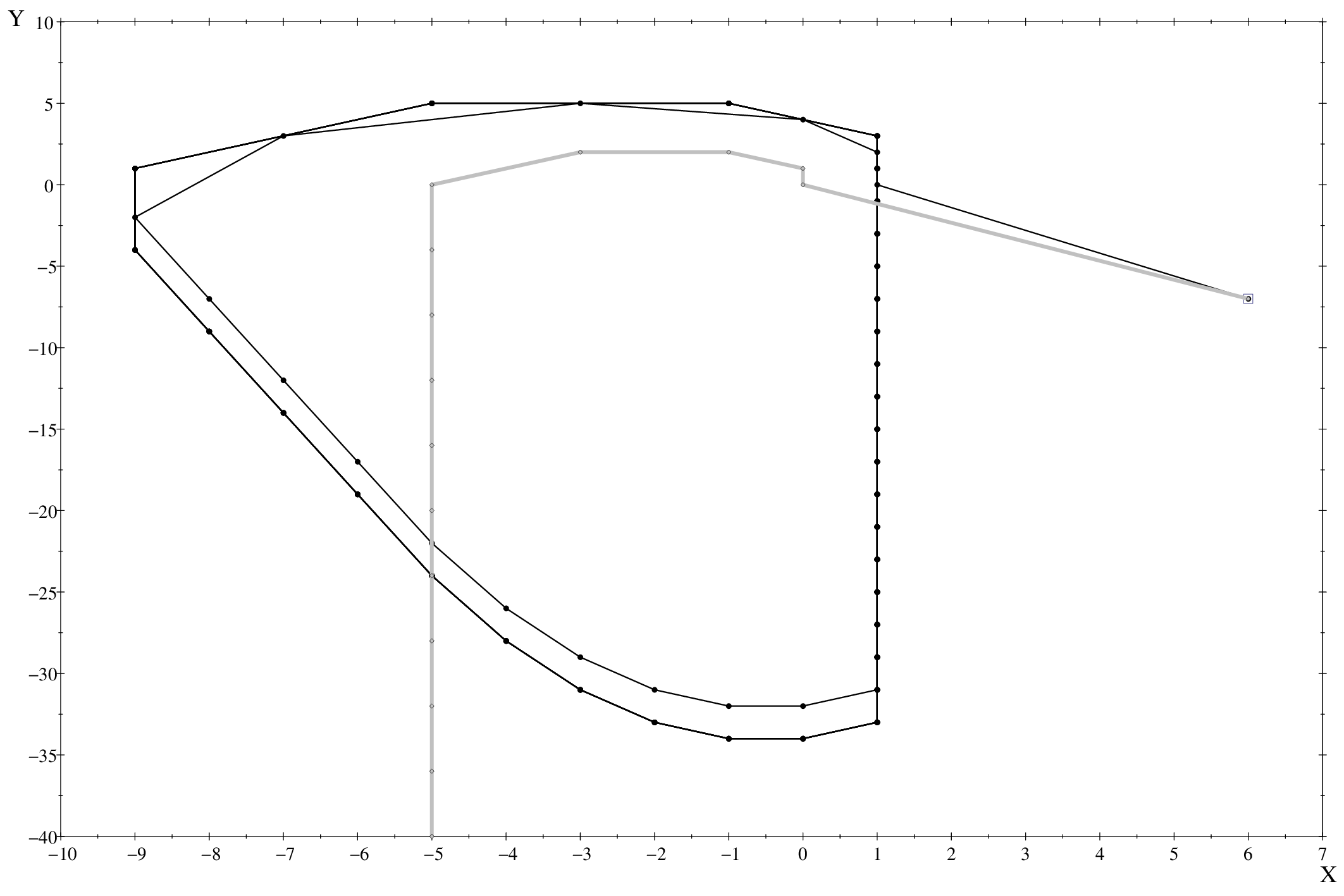}
\caption{Figure showing orbits for the same initial condition $(X_0,Y_0)=(6,-7)$ and parameter values ($\Gamma=6, \Delta=5$), for two different instances of the ultradiscrete system \eqref{LVextud} ($T=\pm1$), the orbit in one of which ($T=+1$) is instantly attracted to a periodic orbit with period 32 (the plot showing the first 150 iterations of the system), whereas that in the other ($T=-1$) is attracted to a fixed point at infinity (the plot showing the first 15 iterations of the system). To enhance visibility, points in the same orbit are connected by lines, black lines for the automaton with a periodic orbit and gray lines for the automaton that only has attracting fixed points.} 
\label{fig2}
\end{center}
\end{figure}

Fig.~\ref{fig2} shows the orbits of $(X_0,Y_0)=(6,-7)$ for both cellular automata, when $\Gamma=6$ and $\Delta=5$, one orbit (for \eqref{LVextud} with $T=1$) hitting a periodic orbit (of period 32) almost immediately, and the other one (for \eqref{LVextud} with $T=-1$)  being attracted to the sink at $(-5,-\infty)$.

Hence we claim that the cellular automaton obtained by mimicking, at the ultradiscrete limit, the {\em continuum limit} $\tau=e^{T/\epsilon}\overset{\epsilon\to+0}{\longrightarrow} 0$,
\begin{equation}
\left\{
\begin{array}{l}
X_{n+1} = X_n - \max[0,X_n,Y_n]\\[-1mm] \\
Y_{n+1}= Y_n+\max[0,\Gamma+X_n]-\max[0,\Delta],
\end{array}
\right.\label{LVextuc}
\end{equation}
has dynamics that are very similar to those of the original continuous system \eqref{LVext} for all values of the parameters---i.e. the same number of fixed points, exactly one of which is attractive depending on the parameters and absence of a limit cycle---whereas the intermediate discrete system used in the construction did not. As far as we know this is the first instance of such a construction.

\section{Conclusion}
\begin{figure}[h]
\begin{center}
\includegraphics[width=11cm]{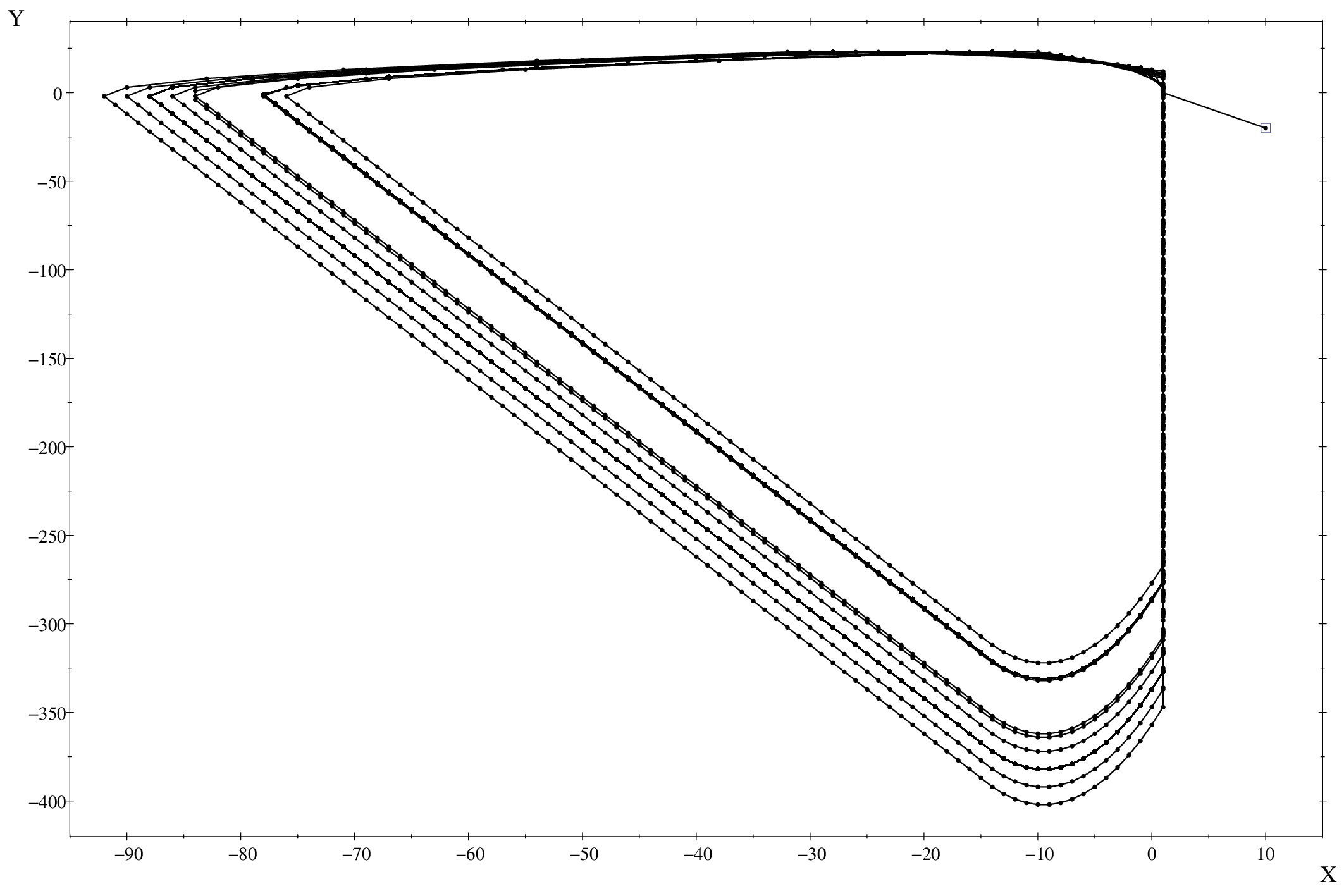}
\caption{Figure showing a complicated attractive periodic orbit for the ultradiscrete system \eqref{LVextud} in the fully discrete regime ($T=1$), the periodicity of which depends on the initial condition in the evolution. The initial point of the orbit is marked by a square box and lines have been added between successive points in the orbit to enhance visibility.} 
\label{fig3}
\end{center}
\end{figure}

The example we gave here is very simple and was only intended as a proof of concept. We believe that many more examples, involving much more complicated dynamics, can be constructed quite easily along the same lines. However, the exact mathematical reasons for this remarkable persistence of dynamics (be it that of the original continuous model or that of the intermediate discrete one) at the ultradiscrete limit, are still very much a mystery. Especially the presence of limit cycles poses several mathematical problems. Very often the  ultradiscretisation procedure results in an automaton on a finite phase space (the exact size of which may be dictated by the values of the initial conditions and parameters). In which case it is not surprising that a limit cycle that exists on the continuous or discrete side shows up as an attracting periodic orbit in the automaton, but how and why this transition occurs is still very much unclear. Fig.~\ref{fig3} shows the periodic orbit that the point $(10,-20)$ is attracted to under the evolution described by \eqref{LVextud} for $T=1, \Gamma=15, \Delta=5$. Varying the initial conditions (for the same parameter values)  even slightly, results in very different periodic orbits which can be vastly longer than the one depicted in Fig.~\ref{fig3}, but also much shorter. Some attempts have been made at explaining this phenomenon [16--18] for specific examples, but as yet there is no rigorous mathematical framework for describing such transitions in general.

\begin{credits}
\subsubsection{\ackname} The author gratefully acknowledges financial support from Arithmer Inc. through a collaborative research grant.

\subsubsection{\discintname}
The author has no competing interests to declare that are
relevant to the content of this article. 
\end{credits}

\end{document}